\documentclass[11pt]{article}

\usepackage[margin=1in]{geometry}
\usepackage{amsmath, amssymb, amsthm}
\usepackage{mathtools}
\usepackage{authblk}
\usepackage[numbers,sort&compress]{natbib}
\usepackage{hyperref}
\usepackage{enumitem}
\usepackage{setspace}
\usepackage{microtype}

\hypersetup{
    colorlinks=true,
    linkcolor=blue,
    citecolor=blue,
    urlcolor=blue
}

\newtheorem{theorem}{Theorem}
\newtheorem{definition}{Definition}

\title{\textbf{Canonical Functionalism:}\\
Defining Functional Structure without Observer-Relative Semantic Maps}

\author[1]{Ryota Kanai}
\author[2,3]{Shuqin Ma}
\affil[1]{Araya Inc., Tokyo, Japan}
\affil[2]{School of Philosophy, Fudan University, Shanghai, China}
\affil[3]{Sussex Centre for Consciousness Science, University of Sussex, Brighton, UK}

\date{}

\begin{document}

\maketitle

\begin{abstract}
Computational functionalism about consciousness is often criticized for relying on observer-relative interpretations of physical systems. This paper proposes a mathematical refinement of functionalism that avoids this problem. The central idea is that consciousness-relevant functional organization should be identified not with arbitrary input-output mappings, semantic labels, or externally imposed computational descriptions, but with a system's \emph{canonical functional structure}: the minimal state-transition structure obtained by identifying internal states that have identical future behavior under all possible continuations.

On this view, a state is functionally defined by its complete counterfactual role: how the system would evolve and respond from that state under possible future interactions. We call this position \emph{canonical functionalism}. The framework does not claim to identify which systems are conscious, nor to show that functional organization is sufficient for consciousness. Rather, it identifies the canonical object over which functionalist theories of consciousness should be formulated: the task is to specify consciousness-relevant invariants, measures, or structural conditions over canonical functional structures, rather than over arbitrary semantic interpretations or superficial behavioral profiles. This reframes familiar objections about lookup tables, simulations, unfolding, and observer-relative computation: such cases do not by themselves refute functionalism, but force the functionalist to specify whether the relevant canonical structure is preserved, and if not, which additional structural features are missing.
\end{abstract}

\section{Introduction}

Functionalism is one of the central positions in the philosophy and science of mind. In its classical form, it individuates mental states by their causal or functional roles rather than by their material constitution \citep{putnam1960minds,lewis1972psychophysical,block1978troubles,chalmers1996conscious}. A mental state is characterized by what typically causes it, how it interacts with other internal states, and what kinds of behavior or further processing it tends to produce. This idea is attractive partly because it supports multiple realizability: if the relevant organization is functional, then the same mental property could in principle be realized by different physical substrates.

The possibility of artificial consciousness has made this issue newly urgent \citep{dehaene2017consciousness,butlin2023consciousness,chalmers2023large,long2024welfare}. If consciousness depends on functional or computational organization, then artificial systems are not excluded in principle. If, however, computation is merely an interpretation imposed on physical dynamics by an observer, then computational functionalism appears unable to ground consciousness as an intrinsic property of a system. This worry has been expressed in several forms: Searle-style arguments against syntax as sufficient for semantics \citep{searle1980minds}, triviality arguments about computational implementation \citep{sprevak2018triviality}, concerns about representation and individuation in computation \citep{sprevak2010computation}, and mechanistic objections to purely abstract accounts of physical computation \citep{piccinini2015physical}. Recent substrate-sensitive challenges, including biological naturalist arguments, similarly question whether computation alone can be sufficient for consciousness \citep{seth2025conscious,block2026meat}. A recent formulation, the ``abstraction fallacy'' objection, argues that artificial systems may simulate consciousness without instantiating it if their computational organization is only an abstract syntactic description rather than an intrinsic physical process \citep{lerchner2026abstraction}.

These objections identify a genuine weakness in naive computational functionalism. Consciousness should not be inferred from behavioral mimicry, linguistic self-reports, or an arbitrary mapping from physical states to mentalistic symbols \citep{searle1980minds,harnad1990symbol}. Nor is it enough to say that two systems implement the same input-output table while ignoring the physical and dynamical organization that realizes their behavior. A theory of consciousness should not depend on semantic labels assigned from outside the system.

The conclusion, however, need not be anti-functionalist. Objections based on observer-relative maps succeed against naive computational functionalism, but they do not show that all computational or functional organization is observer-relative. The relevant distinction is between extrinsic computational interpretation, in which an observer assigns semantic labels to physical states, and intrinsic computational organization, in which a physical system realizes causal, counterfactual, and dynamical structures that constrain its possible futures. The present paper develops this distinction into a formal proposal: functionalist theories of consciousness should be formulated over a canonical structure determined by a system's counterfactual transition organization, rather than over arbitrary semantic mappings imposed from outside.

Semantic labels may be conventional, but counterfactual transition structure is not arbitrary in the same way. The task is therefore to identify a functional structure that is mathematically well defined, physically realized, and not semantically imposed.

In this paper, we use ``intrinsic'' in a specific anti-interpretivist sense. It does not mean independent of all environmental relations. Instead, it means that functional roles are determined through possible interactions, interventions, and counterfactual transitions, not by an observer's arbitrary semantic mapping. Similarly, ``behavior'' refers not to actual observed behavior or a finite behavioral sample, but to the complete counterfactual input-output profile under a specified interface.

We propose that functionalism should take as its primary mathematical object the \emph{canonical functional structure} of an interactive system. This structure is obtained by identifying internal states that have identical future behavior under all admissible future input histories. It is therefore a quotient state-transition system: a structure over equivalence classes of states, equipped with induced dynamics.

Once the relevant interface of inputs, outputs, and interventions is specified, the equivalence relation is fixed by the system's counterfactual organization. The canonical functional structure is thus externally accessible but not externally labeled. It is canonical in the mathematical sense that any two systems with the same complete counterfactual profile determine isomorphic quotient structures.

This gives functionalism a more precise target. A functionalist theory should not merely claim that consciousness depends on ``functional organization'' in an unspecified sense. It should state the conditions under which a canonical functional structure, or an appropriately enriched version of it, gives rise to consciousness. The framework developed below does not prove functionalism; it identifies the kind of object over which a rigorous functionalist theory of consciousness should be formulated.

The proposal also connects with the desideratum of universality in consciousness science \citep{kanai2024universal}. A universal theory should not be parochially tied to human biology, yet it must also avoid the opposite error of treating arbitrary interpretation as sufficient. Canonical functional structures offer one way to express this balance: they are substrate-general in form, but constrained by physically realized counterfactual organization. This makes them suitable candidates for formulating consciousness-relevant principles that could apply across biological, artificial, and other possible systems without collapsing into observer-relative attribution.

The paper proceeds as follows. We first introduce the intuitive idea that a functional state is a structured space of possible futures. We then distinguish the proposal from behaviorism and from observer-relative computational interpretation. The formal sections define the canonical functional structure and prove that systems with the same complete behavior share an isomorphic canonical structure. Subsequent sections discuss the dynamical content of the quotient, its relation to computational mechanics and \(\epsilon\)-machines, its compatibility with universality, and its implications for debates about simulation, instantiation, and artificial consciousness.

\section{The Intuitive Idea: A State Is a Space of Possible Futures}

To understand the proposal, consider a simple elevator controller in an office building.

At a low physical level, the controller may be implemented by relays, a microprocessor, a cloud-connected software system, or a neural-network-based controller. These implementations may be very different. But at the functional level, the elevator has states such as:
\begin{itemize}[leftmargin=2em]
    \item idle at floor 1 with doors closed;
    \item moving upward toward floor 5;
    \item doors open at floor 3;
    \item waiting with a pending request from floor 7;
    \item overloaded and unable to move;
    \item emergency stop mode.
\end{itemize}

These states are not defined by their material composition. They are defined by what the elevator will do next under possible interactions. A state such as ``doors open at floor 3 with a pending request for floor 7'' is the state from which pressing the close-door button may close the doors, pressing the alarm button may enter emergency mode, a timeout may close the doors automatically, and the next movement may be upward rather than downward.

The state is therefore defined by its \emph{future possibilities}.

Now suppose the controller has two different internal microstates. Perhaps one is encoded in one memory register and another in a different register. Or perhaps one is implemented by a relay configuration and the other by software variables. If every possible future interaction produces exactly the same outputs from both states, then functionally they are the same state. Nothing a user, technician, or environment can do within the relevant interface will distinguish them. They have the same role in the system's space of possible interactions.

The canonical functional structure is obtained by collapsing all such indistinguishable states.

This idea generalizes beyond elevators. For any interactive system, an internal state can be understood in terms of the future behavior it makes possible. A state is not merely what the system is doing now. It is a disposition toward a structured family of possible futures.

This point is especially important for consciousness. Functionalism has never been the view that mental states are individuated merely by the behavior that happens to occur at a given moment. A belief, desire, perceptual state, or pain state is individuated by its role within a network of possible causal relations: how it would affect attention, memory, action, report, learning, and further cognition under different possible conditions. The relevant structure is therefore not a finite record of actual responses, but a counterfactual profile of possible transitions.

A serious functionalism about consciousness must therefore be state-based and transition-sensitive. It must characterize the organized space of possible ways in which a system could evolve from a given state under possible inputs, perturbations, and internal conditions. The canonical functional structure gives a mathematical form to this intuition by representing each state through its complete future interaction profile.

This use of behavioral equivalence should not be confused with behaviorism. The relevant object is not the actual behavior a system happens to display, nor a finite set of observed stimulus-response pairs. It is the complete counterfactual profile of possible state transitions and outputs under admissible future interactions. The canonical structure is therefore state-based and transition-sensitive: it captures the organization of possible dynamics from each state, rather than merely cataloguing outward behavior.

\section{From Observer-Relative Maps to Canonical Structures}

A common objection to computational functionalism is that it relies on observer-relative mappings, a concern closely related to debates over computational implementation and mechanistic accounts of physical computation \citep{piccinini2015physical,sprevak2010computation}. An external observer can assign many labels to a physical system. A voltage can be called ``1,'' a neural activation can be called ``red,'' a hidden vector can be called ``cat,'' and a trajectory can be interpreted as a decision process. Such labels are often useful, but they are not intrinsic to the system, a point closely related to the symbol-grounding problem \citep{harnad1990symbol}.

However, the system's pattern of possible state transitions is not arbitrary in the same way. If a physical state leads to different future states under different inputs, if it stores information, if it constrains downstream processes, if it is reachable from some histories and not others, these facts are not created by semantic labeling. They are facts about the system's structure and dynamics.

The canonical functional structure captures exactly this kind of organization. It does not ask what labels an observer assigns to the states. It asks which states are distinguishable by their future roles.

This is why the proposal is not vulnerable to the simplest mapmaker objection. The canonical structure is not produced by assigning meanings to states. It is produced by quotienting the system's state space by an equivalence relation defined over possible future behaviors.

\section{Canonical Functionalism}

We can now state the central philosophical thesis.

\begin{definition}[Canonical functionalism]
Consciousness-relevant functional organization is not an arbitrary computational interpretation imposed on a system, nor merely its actual input-output behavior. It is the system's \emph{canonical functional structure}: the minimal structure of functionally distinguishable states and transitions determined by the system's complete counterfactual behavior.
\end{definition}

This view has several components.

First, mental states are not identified with raw physical microstates. Many physical differences may be irrelevant to consciousness.

Second, mental states are not identified with externally imposed semantic labels. A system is not conscious merely because an observer interprets its states as beliefs, perceptions, or feelings.

Third, mental states are not identified with actual behavior alone. Consciousness depends, if functionalism is correct, on the structure of possible transitions among states.

Fourth, functional identity is defined by counterfactual equivalence. Two states are functionally identical if they cannot be distinguished by any possible future interaction in the relevant domain.

Fifth, the resulting structure is canonical. Once the relevant input-output-intervention interface is fixed, the quotient structure is mathematically determined.

This gives functionalism a more rigorous form. Instead of saying vaguely that ``consciousness depends on functional organization,'' canonical functionalism says:

\begin{quote}
Consciousness depends on properties defined over the canonical functional structure of a system.
\end{quote}

For a conscious property \(C\), the canonical functionalist claims that \(C\) should be invariant under isomorphism of canonical functional structures. That is, if two systems have the same canonical functional structure, then they cannot differ in consciousness-relevant functional properties.

This is not yet a complete theory of consciousness. It does not tell us which canonical structures are conscious. But it tells us what kind of object a functionalist theory of consciousness should study.

\section{The Formal Framework}

We now turn to the mathematical formulation.

Let \(I\) be a set of possible inputs and \(O\) a set of possible outputs. Let \(I^*\) denote the set of finite input histories.

Here, \(w\in I^*\) denotes a finite input history, while \(\epsilon\) denotes the empty history, i.e. the case in which no further input is supplied. For example, if \(w=i_1i_2i_3\), then \(w\) represents the sequence in which the inputs \(i_1\), \(i_2\), and \(i_3\) are applied in that order. The special history \(\epsilon\) has length zero.

We model an interactive deterministic system as a Moore-type machine:
\[
S=(X,x_0,\delta,o),
\]
where \(X\) is the set of internal states, \(x_0\in X\) is the initial state,
\[
\delta:X\times I\to X
\]
is the state-transition function, and
\[
o:X\to O
\]
is the output function.

The transition function extends to finite input histories:
\[
\delta^*(x,\epsilon)=x,
\]
\[
\delta^*(x,wi)=\delta(\delta^*(x,w),i).
\]

For each state \(x\in X\), define its complete future behavior:
\[
b_x^S:I^*\to O
\]
by
\[
b_x^S(w)=o(\delta^*(x,w)).
\]

Thus, \(b_x^S(w)\) is the output obtained by starting from state \(x\) and then applying the future input history \(w\). In particular, \(b_x^S(\epsilon)=o(x)\).

Thus \(b_x^S\) tells us what output the system would produce after every possible future input history \(w\), starting from state \(x\).

The overall behavior of the system from its initial state is:
\[
B_S=b_{x_0}^S.
\]

Now define an equivalence relation on states:
\[
x\sim_S y
\]
if and only if
\[
\forall w\in I^*, \quad b_x^S(w)=b_y^S(w).
\]

In words, two states are equivalent if no possible future input history can distinguish them.

\begin{definition}[Reachable states]
The reachable state set of \(S\) is
\[
R_S=\{\delta^*(x_0,u)\mid u\in I^*\}.
\]
\end{definition}

\begin{definition}[canonical functional structure]
The \emph{canonical functional structure} of \(S\) is the quotient:
\[
\mathrm{Can}(S)=R_S/{\sim_S}.
\]
\end{definition}

The quotient inherits a transition structure:
\[
\bar{\delta}([x],i)=[\delta(x,i)],
\]
and an output function:
\[
\bar{o}([x])=o(x).
\]

These are well-defined because \(\sim_S\) is compatible with both output and transition.

If \(x\sim_S y\), then:
\[
o(x)=b_x^S(\epsilon)=b_y^S(\epsilon)=o(y).
\]

And for any \(i\in I\), if \(x\sim_S y\), then for all \(w\in I^*\):
\[
b_{\delta(x,i)}^S(w)
=
b_x^S(iw)
=
b_y^S(iw)
=
b_{\delta(y,i)}^S(w).
\]
Therefore:
\[
\delta(x,i)\sim_S \delta(y,i).
\]

So the quotient is a well-defined state-transition system.

It is important to emphasize that the canonical functional structure is not merely a static partition of the original state space. The equivalence relation \(\sim_S\) is compatible with the transition dynamics of the system. In algebraic terms, it is a congruence with respect to the input-indexed transition maps. This is why the quotient inherits not only a set of equivalence classes, but a genuine input-driven dynamics.

\section{The Canonical Realization Theorem}

We can now state the core theorem.

\begin{theorem}[Canonical functional structure]

Let
\[
S=(X,x_0,\delta,o)
\]
and
\[
T=(Y,y_0,\gamma,p)
\]
be two deterministic interactive systems over the same input set \(I\) and output set \(O\). If they have the same complete input-output behavior,
\[
B_S=B_T,
\]
then their canonical functional structures are isomorphic:
\[
\mathrm{Can}(S)\cong \mathrm{Can}(T).
\]
Moreover, this common structure is isomorphic to a canonical machine determined uniquely by the behavior \(B:I^*\to O\).
\end{theorem}

\subsection{Construction of the canonical machine}

Given any behavior
\[
B:I^*\to O,
\]
define, for each input history \(u\in I^*\), the residual behavior
\[
B_u:I^*\to O
\]
by
\[
B_u(v)=B(uv).
\]

Let:
\[
Q_B=\{B_u\mid u\in I^*\}.
\]

Define:
\[
q_0=B_\epsilon=B,
\]
\[
\partial(B_u,i)=B_{ui},
\]
and
\[
o_B(B_u)=B_u(\epsilon)=B(u).
\]

Then:
\[
M_B=(Q_B,q_0,\partial,o_B)
\]
is the canonical realization of \(B\).

As noted above, this machine is analogous to minimal-state constructions in automata theory. Its states correspond not to physical microstates, but to residual future behaviors. A state is ``where the system is'' in the space of possible future responses.

Mathematically, this construction is closely related to classical minimal realization and Myhill–Nerode style equivalence for deterministic automata. The novelty of the present proposal is not the quotient construction itself, but its use as a principled target for functionalist theories of consciousness: it identifies the structure over which consciousness-relevant functional properties should be defined if functionalism is to avoid observer-relative semantic mappings.
\subsection{Proof}

First, \(M_B\) realizes \(B\). After reading history \(u\), the machine is in state \(B_u\). Its output is:
\[
o_B(B_u)=B_u(\epsilon)=B(u).
\]
Thus the behavior of \(M_B\) from \(q_0\) is exactly \(B\).

Second, let \(S=(X,x_0,\delta,o)\) be any system realizing \(B\). We consider its reachable part \(R_S\), since unreachable states do not contribute to the behavior from the initial state.

Define:
\[
h_S:R_S\to Q_B
\]
by
\[
h_S(\delta^*(x_0,u))=B_u.
\]

This is well-defined. If:
\[
\delta^*(x_0,u)=\delta^*(x_0,v),
\]
then for all \(z\in I^*\):
\[
B(uz)=o(\delta^*(x_0,uz))
\]
\[
=o(\delta^*(\delta^*(x_0,u),z))
\]
\[
=o(\delta^*(\delta^*(x_0,v),z))
\]
\[
=B(vz).
\]
Therefore:
\[
B_u=B_v.
\]

The map \(h_S\) preserves outputs:
\[
o_B(h_S(\delta^*(x_0,u)))
=
o_B(B_u)
=
B(u)
=
o(\delta^*(x_0,u)).
\]

It also preserves transitions:
\[
h_S(\delta(\delta^*(x_0,u),i))
=
h_S(\delta^*(x_0,ui))
=
B_{ui}
=
\partial(B_u,i)
=
\partial(h_S(\delta^*(x_0,u)),i).
\]

Finally, the kernel of \(h_S\) is exactly behavioral equivalence:
\[
h_S(x)=h_S(y)
\]
if and only if
\[
b_x^S=b_y^S.
\]

Therefore:
\[
R_S/{\sim_S}\cong Q_B.
\]

The same argument applies to \(T\). Hence:
\[
\mathrm{Can}(S)\cong Q_B\cong \mathrm{Can}(T).
\]

So any two systems with the same complete behavior share the same canonical functional structure.
\qed

The requirement of complete future behavioral equivalence should be understood as an idealization. Empirical applications will require approximate, probabilistic, finite-horizon, or coarse-grained versions of the canonical quotient. The ideal construction is nevertheless useful because it specifies the limiting object that such approximations attempt to estimate.

\subsection{A simple example}

A simple example may help clarify what the canonical quotient preserves. Let
\[
I=\{0,1\},\qquad O=\{\mathrm{even},\mathrm{odd}\}.
\]
Consider a system with four internal states,
\[
X=\{a,b,c,d\},
\]
with outputs
\[
o(a)=o(b)=\mathrm{even},\qquad o(c)=o(d)=\mathrm{odd}.
\]
The transition function is given by
\[
\begin{array}{c|c|c|c}
\text{state} & o(x) & \delta(x,0) & \delta(x,1)\\
\hline
a & \mathrm{even} & b & c\\
b & \mathrm{even} & a & d\\
c & \mathrm{odd} & d & a\\
d & \mathrm{odd} & c & b
\end{array}
\]
The input \(0\) preserves parity, while the input \(1\) reverses parity. The distinction between \(a\) and \(b\), and likewise between \(c\) and \(d\), is an implementation-level distinction that never affects future outputs. For every future input history \(w\in\{0,1\}^*\),
\[
b_a^S(w)=b_b^S(w),
\]
and
\[
b_c^S(w)=b_d^S(w).
\]
Thus,
\[
a\sim_S b,\qquad c\sim_S d.
\]
However, \(a\not\sim_S c\), since
\[
b_a^S(\epsilon)=\mathrm{even}
\]
whereas
\[
b_c^S(\epsilon)=\mathrm{odd}.
\]

The canonical quotient therefore has two states:
\[
E=[a]=\{a,b\},\qquad O=[c]=\{c,d\}.
\]
The induced output and transition functions are
\[
\bar{o}(E)=\mathrm{even},\qquad \bar{o}(O)=\mathrm{odd},
\]
and
\[
\bar{\delta}(E,0)=E,\qquad \bar{\delta}(E,1)=O,
\]
\[
\bar{\delta}(O,0)=O,\qquad \bar{\delta}(O,1)=E.
\]
Equivalently,
\[
\begin{array}{c|c|c|c}
\text{canonical state} & \bar{o} & \bar{\delta}(\cdot,0) & \bar{\delta}(\cdot,1)\\
\hline
E & \mathrm{even} & E & O\\
O & \mathrm{odd} & O & E
\end{array}
\]

This example illustrates that the canonical functional structure is not merely a partition of the original state space. It is a reduced dynamical system. The quotient removes implementation-level distinctions that make no difference to any possible future interaction, while preserving the input-driven transitions that determine future behavior.

\section{Philosophical Implications of the Theorem}

The canonical realization theorem has a direct philosophical consequence. It shows that the functional structure relevant to functionalism need not remain an informal or metaphorical notion. Once an interface of possible inputs and outputs is specified, there exists a mathematically well-defined object associated with any interactive system: its canonical functional structure. This structure is obtained by quotienting the reachable state space by complete future behavioral equivalence. It is therefore not an arbitrary description imposed on the system, but a determinate structure generated by the system's own pattern of counterfactual transitions.

This point is important for the interpretation of functionalism. Functionalism is often stated as the claim that mental states are individuated by their causal or functional roles. However, without a precise account of what such roles are, the thesis remains vulnerable to two opposite objections. On the one hand, it may collapse into a superficial input-output behaviorism. On the other hand, it may appear to depend on observer-relative semantic interpretation, where an external theorist assigns mental or computational labels to physical states. The present framework avoids both problems. The canonical functional structure is richer than actual input-output behavior, because it encodes the full space of possible future interactions. At the same time, it does not depend on externally assigned semantic labels, because its states are equivalence classes determined by future behavioral roles.

Canonical functionalism is continuous with, but more restrictive than, classical functionalism. Putnam's early functionalism treated mental states as defined by their causal roles \citep{putnam1960minds}, but his later critique in \emph{Representation and Reality} emphasized the instability of functional and representational descriptions when they depend on externally imposed interpretations \citep{putnam1988representation}. The present framework accepts this worry as genuine. Its response is not to abandon functionalism, but to replace observer-relative functional maps with a canonical quotient fixed by counterfactual transition structure. In this sense, canonical functionalism aims to recover the anti-chauvinist motivation of functionalism while avoiding the indeterminacy that motivated Putnam's later dissatisfaction with it.

The proposal also sharpens Chalmers's principle of organizational invariance. Chalmers argues that systems with the same fine-grained functional organization should have the same conscious experience \citep{chalmers1996conscious}. Canonical functionalism provides a formal candidate for the relevant notion of organization: two systems are consciousness-equivalent, insofar as functionalism is correct, when their consciousness-relevant properties factor through isomorphic canonical functional structures. The claim is therefore not merely that consciousness is preserved under ``same organization'' in an intuitive sense, but that the relevant organization is the canonical structure of future-directed functional roles.

Finally, the framework responds to Block-style worries about overly liberal functionalism. Block's objections show that functionalism becomes implausible if it identifies mentality with superficial input-output equivalence or with an arbitrary realization of a computation \citep{block1978troubles}. Canonical functionalism accepts this diagnosis but draws a different conclusion: the relevant functional structure must be state-based, counterfactual, transition-sensitive, and canonical. Thus, the theory is not committed to treating every behavioral duplicate or interpreted computation as conscious; it is committed only to the invariance of consciousness-relevant properties under preservation of the appropriate canonical structure.

Thus, the theorem suggests a more precise formulation of functionalism. A functionalist theory of consciousness should not merely claim that consciousness depends on ``functional organization'' in an unspecified sense. It should specify the conditions under which a canonical functional structure gives rise to consciousness. In other words, if consciousness is to be explained functionally, then the explanandum should be connected to properties of
\[
\mathrm{Can}(S),
\]
rather than to raw physical microstates, arbitrary computational descriptions, or isolated input-output mappings.

This also clarifies the sense in which the proposed structure is intrinsic. It is not intrinsic in the sense of being independent of all relations to an environment. On the contrary, it is defined through the system's possible interactions with inputs and outputs. Rather, it is intrinsic in the more relevant anti-interpretivist sense: once the interaction interface is fixed, the canonical functional structure is determined by the system's own transition organization and not by an external observer's semantic labeling. A theorist may choose a level of description or an interface, but she does not freely assign the resulting equivalence classes. They are fixed by the system's counterfactual behavioral organization.

This distinction is central. The canonical structure is externally accessible but not externally imposed. It is accessible because it is defined in terms of possible interactions, interventions, and observable consequences. Yet it is not imposed because the equivalence relation is generated by what the system would actually do under those possible interactions. In this respect, canonical functionalism occupies a middle position between naive behaviorism and hidden-substrate essentialism. It grounds functional organization in observable counterfactual behavior while preserving an internal state-transition structure that is uniquely determined up to isomorphism.

The choice of interface is therefore not a merely technical detail. Different choices of inputs, outputs, interventions, temporal grain, and environmental coupling may yield different canonical structures. This does not make the construction arbitrary; rather, it identifies where substantive theoretical commitments enter. A theory of consciousness must specify which interactions and interventions are relevant for individuating consciousness-supporting functional states.

The philosophical burden for functionalist theories of consciousness can therefore be restated more rigorously. Such theories must identify consciousness-relevant invariants, measures, or structural conditions on canonical functional structures. For example, they may claim that consciousness requires particular forms of recurrent organization, global availability, temporal continuity, self-modeling, integration, closure, or representational geometry. But these claims should be formulated as properties of the canonical or appropriately enriched functional structure, rather than as informal appeals to function or as semantic attributions made by an observer.

The theorem does not by itself establish that functionalism is true. It does not prove that any particular canonical functional structure is conscious. Its significance is more foundational. It identifies the mathematical object that a rigorous functionalist theory of consciousness should take as its target. If consciousness is functionally constituted, then its conditions of occurrence should be stated over canonical functional structures. This provides a disciplined way to formulate, compare, and test functionalist theories while avoiding the charge that computation or function is merely an observer-relative map.

\section{Applications to Familiar Objections and Test Cases}

The canonical framework is not intended merely as an abstract formalism. It also clarifies several familiar debates in the philosophy and science of consciousness. The unfolding argument, lookup-table objections, and the Chinese Room argument can all be understood as test cases for the same underlying question: whether a system that differs in implementation nevertheless realizes the same consciousness-relevant functional structure. The present framework does not settle all such cases by itself, but it states the functionalist position with greater precision.

\subsection{The unfolding argument}

The unfolding argument challenges theories that identify consciousness with recurrent causal structure \citep{doerig2019unfolding}. Very roughly, the argument says that for certain recurrent systems, one can construct feedforward or unfolded systems that reproduce the same input-output behavior. If consciousness depends on recurrence itself, then recurrent and unfolded systems should differ in consciousness. But if they are behaviorally indistinguishable, this creates pressure on such theories.

Canonical functionalism offers a different diagnosis. If a recurrent system \(R\) and an unfolded system \(F\) have the same complete future behavior over the relevant interface, then
\[
\mathrm{Can}(R)\cong \mathrm{Can}(F).
\]
At the level of canonical functional structure, they are therefore the same. Their low-level causal graphs may differ: one may contain explicit cycles, whereas the other may encode the corresponding temporal dependencies in an unfolded architecture. But if the complete counterfactual transition structure is preserved, then the difference is not a functional difference at the canonical level.

This does not show that recurrence is irrelevant to consciousness. Rather, it clarifies what a functionalist should mean by the relevance of recurrence. Recurrence matters insofar as it contributes to canonical functional structure: memory, temporal continuity, self-conditioning, context sensitivity, integration, and future-directed counterfactual organization. If an unfolded system preserves all of these features at the relevant level, canonical functionalism treats it as functionally equivalent. If it does not, then the equivalence fails. Thus, the unfolding argument is best understood not as a refutation of functionalism, but as a demand for precision about which structure the functionalist takes to be consciousness-relevant.

\subsection{Lookup tables}

A second test case is the lookup-table objection. Suppose a gigantic table stores the correct output for every possible input history. Would such a table be conscious if it reproduced the behavior of a conscious system? The force of the objection is that functional equivalence appears to be too easily achieved: if the same outputs can be generated by a sufficiently large table, then functionalism seems committed to attributing consciousness to systems that appear to lack genuine understanding, dynamics, or mentality.

Canonical functionalism distinguishes two cases. A simple table that maps current inputs to current outputs will not generally reproduce the canonical functional structure of a conscious system. It will typically lack the same state transitions, counterfactual dependencies, memory dynamics, sensitivity to history, and internal organization. Such a table may mimic a restricted behavioral profile, but it does not thereby instantiate the same canonical structure.

However, a sufficiently expanded lookup table could, in principle, encode every possible residual behavior \(B_u\). Such a table would contain, explicitly or implicitly, a state corresponding to each possible history and a rule specifying the transition to the next residual behavior for every possible input. In that limiting case, it would realize a structure isomorphic to the canonical machine \(M_B\). The relevant mathematical point is not that the table resembles a brain at the implementation level, but that the quotient structure of functional states and transitions can be the same:
\[
\mathrm{Can}(S)\cong M_B.
\]

This result makes the functionalist position explicit. If a functionalist theory identifies the consciousness-relevant reality with the canonical functional structure, then a system that truly realizes an isomorphic structure cannot be rejected merely because its low-level implementation is a table or an unfamiliar physical mechanism. From within such a functionalist framework, what matters is not whether the implementation appears intuitively mental, but whether it instantiates the relevant canonical organization. If it does, then denying the corresponding conscious property requires adding a further consciousness-relevant condition beyond canonical functional organization.

This should not be read as the author's endorsement of the claim that lookup tables are conscious. The point is conditional and diagnostic. Canonical functionalism identifies what a committed functionalist must say: if consciousness is wholly determined by the appropriate canonical functional structure, then implementation-level differences that preserve that structure are not, by themselves, grounds for denying consciousness. Conversely, if one thinks that a lookup table still lacks consciousness despite realizing the same canonical functional structure, then one must specify which additional property is missing.

\subsection{The Chinese Room}

Searle's Chinese Room argument raises a closely related challenge \citep{searle1980minds}. In that argument, a person who does not understand Chinese follows a rulebook for manipulating Chinese symbols and produces outputs that are indistinguishable from those of a competent Chinese speaker. The intended conclusion is that syntactic symbol manipulation, even if behaviorally successful, is not sufficient for understanding.

From the perspective of canonical functionalism, the Chinese Room is not merely a case about syntax or symbol manipulation. It is a case about whether the room as a whole realizes the relevant canonical functional structure. If the room merely produces correct answers for a limited set of exchanges, then it may at most mimic a restricted behavioral profile. But if the room, including the person, rulebook, memory, update procedures, and interaction history, realizes the complete counterfactual organization of a Chinese-understanding system, then the functionalist cannot dismiss it simply by pointing to the unusual implementation.

The conclusion is again conditional. Canonical functionalism does not require the author to assert that the Chinese Room understands. Rather, it clarifies the dialectical burden. If a system realizes a canonical functional structure isomorphic to that of a genuinely understanding system, then a functionalist theory must treat it as possessing the same functional property. If one denies this, one must identify a further consciousness- or understanding-relevant property not captured by canonical functional organization. Such a property might involve intrinsic causal power, biological embodiment, temporal continuity, semantic grounding, integrated dynamics, or some other constraint. But it must then be introduced as an explicit addition to the theory, rather than left as an intuition that the implementation is ``merely'' a rulebook.

This is not intended as a decisive refutation of Searle's argument. Rather, it clarifies the dialectical position. If one accepts canonical functionalism, then the absence of ordinary biological or semantic features in the Chinese Room is not by itself sufficient to exclude understanding or consciousness, provided the relevant canonical functional structure is preserved. If one rejects this conclusion, one must identify which additional non-functional or enriched-functional property is missing.

Canonical functionalism therefore does not make lookup-table or Chinese Room worries disappear. It transforms them into precise questions about the target structure. Does the system instantiate the same canonical functional structure under the consciousness-relevant interface, or does it merely approximate a restricted input-output profile? This is a more disciplined formulation than the general claim that the system is a ``mere simulation.'' A system is a mere simulation only relative to a specified account of what would count as instantiating the relevant structure. Canonical functionalism requires that this account be made explicit.

\section{Relation to Universality-Based Approaches}

Canonical functionalism is also naturally aligned with a universality-based approach to consciousness \citep{kanai2024universal}. The motivation behind universality is that a theory of consciousness should not be tied prematurely to one familiar biological implementation. If consciousness is a natural phenomenon, the relevant principles should be capable, at least in principle, of applying across humans, non-human animals, artificial systems, and other possible physical systems.

However, universality creates a danger. A theory can become too permissive if it treats any externally imposed interpretation as sufficient. If a system counts as conscious merely because an observer can map its physical states onto a conscious-like computation, then universality degenerates into observer-relative projection.

Canonical functionalism offers a way to preserve universality without collapsing into arbitrariness. It identifies the relevant object not with semantic labels, biological materials, or anthropocentric behavioral cues, but with a mathematically defined structure of counterfactual state transitions. This structure can be compared across different substrates because it is not specified in biological vocabulary. At the same time, it is not arbitrary because, once the interface is fixed, it is determined by the system's complete future behavior.

In this sense, canonical functionalism supports a constrained form of substrate generality. The claim is not that substrate never matters. Rather, substrate matters insofar as it realizes, stabilizes, or transforms the canonical functional structure relevant to consciousness. Biological neurons, silicon circuits, neuromorphic hardware, organoids, or hybrid brain--machine systems may differ greatly at the physical level, but the functionalist question becomes whether they instantiate the relevant canonical organization.

This clarifies the relation between universality and intrinsicality. A universal theory of consciousness must apply across substrates, but it cannot depend on observer-relative interpretation. The canonical functional structure provides such a target: it is abstract enough to be shared by different implementations, yet fixed by the system's own counterfactual transition organization rather than by arbitrary semantic labeling.

Universality therefore does not mean that consciousness can be assigned to any system by interpretation. It means that consciousness-relevant principles should be formulated at the right level of abstraction: not as parochial biological details, and not as external symbolic maps, but as mathematically characterizable structures realized by physical systems. Canonical functionalism does not itself specify which structures are consciousness-making; it provides the domain over which more specific theories may define conditions such as global availability, predictive control, intrinsic causal power, temporal continuity, or representational geometry.

\section{Relation to Computational Mechanics and \texorpdfstring{\(\epsilon\)}{epsilon}-Machines}

The canonical functional structure proposed here is closely related to the causal-state construction in computational mechanics. In computational mechanics, histories are grouped into causal states when they give rise to the same conditional distribution over futures. The resulting \(\epsilon\)-machine is the minimal predictive model of a stochastic process, unique up to isomorphism under the relevant assumptions \citep{shalizi2001computational}.

The analogy is direct. In the present framework, internal states are grouped together when they give rise to the same complete future input-output behavior. Thus, both constructions define states by their future-directed roles rather than by their intrinsic labels or microscopic implementation. A causal state is a predictive equivalence class of pasts; a canonical functional state is a counterfactual equivalence class of internal states.

There are, however, important differences. First, \(\epsilon\)-machines are primarily predictive structures for stochastic processes, whereas canonical functional structures are introduced here as functional structures for interactive systems. The relevant equivalence relation is not merely sameness of future distribution given a past, but sameness of future behavior under all admissible future inputs, interventions, or interaction histories. Second, computational mechanics typically constructs states from observed histories, whereas the present framework begins with an implemented system and quotients its internal state space by behavioral equivalence. Third, the philosophical role is different: the canonical quotient is proposed as the formal object over which consciousness-relevant functional properties may be defined.

Thus, canonical functionalism can be seen as a consciousness-theoretic generalization of the causal-state idea. It preserves the key insight of computational mechanics---that a state is defined by its consequences for possible futures---but embeds this insight within a functionalist account of consciousness, implementation, and multiple realizability.

\section{Consciousness as a Property over Canonical Structures}

We can now define the core supervenience claim.

Let \(C(S,x)\) denote the consciousness-relevant property of system \(S\) in state \(x\). This may include whether the state is conscious, the level of consciousness, or the quality or content of experience.

Canonical functionalism claims that if:
\[
b_x^S=b_y^T,
\]
then:
\[
C(S,x)=C(T,y).
\]

In other words, if two states have identical complete future behavior under the relevant interface, then they share the same consciousness-relevant functional property.

Equivalently, \(C\) factors through the canonical quotient:
\[
C(S,x)=\widehat{C}([x]).
\]

Here \([x]\) is the equivalence class of \(x\) in the canonical functional structure. The hat notation indicates that \(\widehat{C}\) is defined on the quotient structure. While \(C(S,x)\) assigns a consciousness-relevant property to a particular system-state pair, \(\widehat{C}([x])\) assigns that property to the canonical functional state represented by the equivalence class \([x]\). Thus, canonical functionalism requires \(C\) to factor through the quotient map \(x\mapsto [x]\).

This is the formal heart of canonical functionalism.

It says that consciousness-relevant functional properties are not properties of raw physical states \(x\), but of their equivalence classes under complete future behavioral role.

This does not mean that physical implementation is irrelevant. The physical system must instantiate the transitions. But once instantiated, consciousness-relevant functional properties are invariant under canonical functional equivalence.

\section{Scope and Theoretical Commitments}

Canonical functionalism is not intended as a complete theory of consciousness. Its primary aim is to identify the formal object over which a functionalist theory of consciousness should be stated. It does not by itself determine which systems are conscious, nor does it specify a sufficient condition for consciousness. Instead, it constrains the form that such a condition should take: if consciousness is functionally constituted, then consciousness-relevant properties should be formulated over canonical functional structures, or over appropriately enriched versions of them.

This distinction matters for artificial consciousness. The present framework does not imply that current AI systems are conscious merely because they display sophisticated linguistic or behavioral capacities. A system's verbal reports, task performance, or apparent self-description are not sufficient by themselves. What matters, on the present view, is whether the system realizes the relevant canonical counterfactual organization. Determining whether any existing AI system does so requires substantive empirical and theoretical analysis.

Nor does canonical functionalism claim that arbitrary computational interpretation is sufficient for consciousness. A physical system does not acquire consciousness because an observer assigns mentalistic labels to its states. The canonical structure is not a semantic map imposed from outside; it is defined by the system's pattern of possible transitions under a specified interface. This is precisely why the framework is intended to avoid the trivialization problems that arise for overly permissive accounts of computational implementation.

At the same time, the view does not treat low-level physical implementation as irrelevant. Physical systems must instantiate the relevant transitions, and different physical details may matter if they affect the consciousness-relevant structure. If temporal scale, causal integration, embodiment, metabolism, stochasticity, or internal intervention-responsiveness is relevant to consciousness, then these features should be incorporated into the interface or into an enriched version of the canonical structure. The framework is therefore not a denial of physical realization, but a way of specifying which physical differences make a functional difference.

The construction is also idealized. Complete future behavioral equivalence is not directly available in empirical practice. Real applications will require approximate, probabilistic, finite-horizon, and coarse-grained versions of the canonical quotient. The ideal formulation is nevertheless useful because it identifies the limiting structure that such approximations attempt to capture.

The contribution of canonical functionalism is therefore foundational rather than diagnostic. It does not answer, by itself, which systems are conscious. It clarifies what a functionalist answer should be about. The central question becomes: which properties of canonical functional structures are necessary or sufficient for consciousness?

\section{What Canonical Functionalism Clarifies}

Canonical functionalism clarifies several issues that have often remained implicit in debates about functionalism and consciousness.

First, it provides a precise mathematical object for functionalist theorizing. Traditional functionalism claims that mental states are individuated by their causal or functional roles, but the relevant notion of a functional role is often left informal. The present framework proposes that, at least for interactive systems of the kind considered here, the relevant object is the canonical functional structure \(\mathrm{Can}(S)\). This structure is not a metaphorical description of function, but a formally defined state-transition system obtained by identifying states with identical complete future behavior. A functionalist theory of consciousness can therefore be formulated as a theory of which properties of \(\mathrm{Can}(S)\), or of an appropriately enriched version of it, are sufficient or necessary for consciousness.

Second, canonical functionalism distinguishes functionalism from simple behaviorism. The canonical structure is defined through possible input-output interactions, but it is not exhausted by actual behavior. It encodes the system's counterfactual organization: how its states would evolve under possible future inputs, interventions, and histories. This is important because mental-state individuation has always depended on more than observed responses. A belief, perception, pain state, or conscious episode is characterized by how it would interact with memory, attention, action, report, learning, and other internal states under a range of possible conditions. The canonical quotient captures this future-directed role structure rather than merely recording a finite behavioral profile.

Third, the framework helps answer the charge that computational or functional descriptions are merely observer-relative. The canonical functional structure does not arise from assigning semantic labels to physical states. It is generated by an equivalence relation over possible future behavior. Once the relevant interface of inputs, outputs, and admissible interactions is fixed, the quotient is determined by the system's own transition organization. In this sense, the structure is externally accessible but not externally imposed. It is accessible through possible interaction with the system, yet it is not created by an observer's interpretive mapping.

Fourth, canonical functionalism makes explicit what must be added if one thinks that ordinary functional equivalence is insufficient for consciousness. If temporal continuity, causal integration, embodiment, metabolic self-maintenance, intervention-responsiveness, or representational geometry is consciousness-relevant, then the relevant interface or state-transition model should be enriched so that those differences are represented. The framework does not deny such properties. Rather, it requires theorists to state clearly whether and how those properties enter the consciousness-relevant structure. This turns otherwise vague disputes about ``mere simulation'' or ``real implementation'' into questions about which structural distinctions should be preserved.

Fifth, canonical functionalism clarifies the status of familiar thought experiments. In cases such as unfolding arguments, lookup tables, or the Chinese Room, the central question is not whether two systems appear similar under a superficial description. The question is whether they instantiate isomorphic canonical functional structures under the consciousness-relevant interface. If they do, then a functionalist has principled reason to treat them as equivalent with respect to functional consciousness-relevant properties. If they do not, then the basis of the difference must be located in a specific structural feature that the simplified comparison failed to preserve.

Finally, the framework provides a bridge between multiple realizability and intrinsic organization. It preserves the functionalist idea that the same consciousness-relevant organization may be realized in different substrates. At the same time, it avoids the overly permissive view that any externally imposed computational interpretation is sufficient. What matters is not arbitrary description, but the physically realized pattern of counterfactual transitions that gives rise to a canonical structure. In this respect, canonical functionalism supports the broader motivation of moving from observer-relative computational maps to observer-independent functional organization. It is continuous with the idea that consciousness research should identify intrinsic causal, counterfactual, dynamical, and possibly geometric structures rather than rely on semantic interpretation alone.

\section{Conclusion}

Functionalism about consciousness needs a more precise mathematical foundation. Naive versions are vulnerable to serious objections: observer-relative interpretation, superficial input-output equivalence, lookup table worries, simulation-versus-instantiation arguments, and unfolding challenges.

This paper has proposed \emph{canonical functionalism}. The central idea is that a functionalist theory of consciousness should formulate its consciousness-relevant conditions over a system's \emph{canonical functional structure}: the minimal state-transition structure obtained by identifying states with identical complete future behavior under a specified interface.

This structure is mathematically well-defined. It is the quotient of a system's reachable state space by future-behavior equivalence. It is equivalent to the canonical machine determined by the system's complete counterfactual input-output profile. It captures not merely actual behavior, but the full organization of possible future interactions. At the same time, its empirical application requires substantive choices about interface, temporal grain, admissible interventions, and approximation.

The philosophical significance is that functionalism need not rely on arbitrary semantic maps. A system's canonical functional structure is not created by an observer's interpretation. Given a specified interface, it is fixed by the system's possible transitions.

This gives functionalism a stronger and more intrinsic form. Consciousness, if functional, should be understood as depending not on raw physical substrate, not on external labels, and not on superficial behavior, but on canonical counterfactual organization physically realized by the system.

The deepest question for functionalists remains open:

\begin{quote}
Which canonical functional structures are conscious?
\end{quote}

This question moves the debate beyond the choice between naive computational functionalism and biological naturalism \citep{seth2025conscious,block2026meat}, toward a universal theory of consciousness grounded in observer-independent functional structure.

\bibliographystyle{plainnat}
\bibliography{canonical_functionalism_revised}

\end{document}